\documentclass[aps,preprint]{revtex4}
\usepackage{amssymb}
\usepackage{amsmath}

\input{tcilatex}

\begin{document}

\title{Calculation of Zero-norm States and Reduction of Stringy Scattering
Amplitudes}
\author{Jen-Chi Lee}
\email{jcclee@cc.nctu.edu.tw}
\affiliation{Department of Electrophysics, National Chiao-Tung University, Hsinchu,
Taiwan, R.O.C. \ \ }
\date{\today }

\begin{abstract}
We give a simplified method to generate two types of zero-norm states in the
old covariant first quantized (OCFQ) spectrum of open bosonic string.
Zero-norm states up to the fourth massive level and general formulas of some
zero-norm tensor states at arbitrary mass levels are calculated. On-shell
Ward identities generated by zero-norm states and the \textit{factorization }%
property of stringy vertex operators can then be used to argue that the
string-tree scattering amplitudes of the degenerate lower spin propagating
states are fixed by those of higher spin propagating states at each fixed
mass level. This decoupling phenomenon is, in contrast to Gross's
high-energy symmetries, valid to \emph{all} energies. As examples, we
explicitly demonstrate this stringy phenomenon up to the fourth massive
level (spin-five), which justifies the calculation of two other previous
approaches based on the massive worldsheet sigma-model and Witten's string
field theory (WSFT).
\end{abstract}

\maketitle

\section{Introduction}

The theory of string, as a consistent quantum theory, has no free parameter
and an infinite number of states. It is thus conceivable that there exists
huge hidden symmetry group which is responsible for the ultraviolet
finiteness of the theory. In fact, it was conjectured by Gross \cite{1} more
than a decade ago that an infinite broken gauge symmetries get restored at
energy much higher than the Planck energy. Moreover, he conjectured that,
for the closed string, there existed an infinite number of linear relations
among the scattering amplitudes of different string states that are valid
order by order and are of the \emph{identical form} in string perturbation
theory as $\alpha ^{\prime }$\ goes to infinity. As a result, the scattering
amplitudes of all string states can be expressed in terms of, say, the
dilaton amplitudes. A similar result was presented in Ref \cite{2} for the
open string case.

Soon after, it was discovered that \cite{3} the equations of motion for
massive background fields of the degenerate positive-norm propagating states
can be expressed in terms of those of higher spin propagating states at each
fixed mass level. This decoupling phenomenon was argued to be arisen from
the existence of two types of zero-norm states with the same Young
representations as those of the degenerate positive-norm states in the OCFQ
spectrum. This was demonstrated by using massive worldsheet sigma-model
approach in the lowest order weak field approximation but valid to all
orders in $\alpha ^{\prime }$\ , and thus was, in contrast to Gross's
result, valid to \emph{all} energies. To compare with the usual sigma-model
loop ($\alpha ^{\prime }$) approximation, this result was argued to be a
sigma-model n+1 loop result for the n-th massive level (spin-n+1) \cite%
{3,4,5}. This calculation applies to both open and closed string cases. In a
recent paper \cite{6}, the same decoupling phenomenon was demonstrated by
using WSFT for the open string case up to the spin-five level. It was shown
that the background fields of these degenerate positive-norm states can be
gauged to the higher \textit{rank} fields at the same mass level.

In this paper, we will derive this interesting stringy decoupling phenomenon
from the third and a more direct method, namely, the S-matrix approach. The
key was to explicitly calculate both types of zero-norm states \cite{7} in
the OCFQ spectrum. An infinite number of \textit{nonlinear} relations
between string scattering amplitudes of different string states with the
same momenta at each fixed mass level can then be written down \cite{8}. By
nonlinearity, one means that the coefficients among scattering amplitudes of
different string states depend on the center of mass scattering angle $\phi
_{CM}$ through the dependence of momentum $k$ \cite{9}. These relations, or
stringy on-shell Ward identities are, as in Gross's case, valid order by
order and are of the identical form in string perturbation theory since
zero-norm states should be decoupled from the string amplitudes at each
order of string perturbation theory. These Ward identities, together with
the \textit{factorization }property of stringy vertex operators, will be
used in this paper to express the scattering amplitudes of the degenerate
lower spin propagating states in terms of those of higher spin propagating
states, and thus reduce the number of independent scattering amplitudes at
each fixed mass level. These Ward identities and the resulting decoupling
phenomenon are, in contrast to Gross's high-energy symmetries, valid to 
\emph{all} energies. However, these nonlinear Ward identities, which are
valid to all energies, are difficult to solve. The high-energy limit of
these stringy Ward identities are recently \cite{9} used to explicitly prove
Gross's conjecture on \textit{linear} relations among high-energy scattering
amplitudes of different string states with the same momenta. It was shown
that these stringy Ward identities get simplied as $\alpha ^{\prime
}\rightarrow \infty $, and the number of independent scattering amplitudes
reduces further. As a result, there is only one independent component of
high energy scattering amplitude at each fixed mass level. All other
components of high energy scattering amplitudes are proportional to it.
Moreover, the proportionality constants between scattering amplitudes of
different string states are calculated. These proportionality constants \
were found to be independent of the scattering angle $\phi _{CM}$ and the
loop order $\chi $ of string perturbation theory as conjectured by Gross 
\cite{1,2}. For the case of string-tree amplitudes, a general formula can
even be given \cite{9} to determine all high energy stringy scattering
amplitudes for arbitrary mass levels in terms of those of tachyons - another
conjecture by Gross \cite{1}.

It is now clear that zero-norm states are of crucial importance to uncover
the fundamental symmetries of string theory \cite{9}. The power of zero-norm
states and their direct relation to spacetime $w_{\infty }$ symmetry and
Ward identities \cite{10} of toy 2D string model were stressed in Ref \cite%
{11}. A general formula of 2D zero-norm states at an arbitrary mass levels
with Polyakov's momentum was given in terms of Schur Polynomials. These
zero-norm states were shown to carry the charges of $w_{\infty }$ symmetry,
which was used to determine the tachyon scattering amplitudes \textit{without%
} any integration. In section II of this paper, with the help of a
simplified method to construct D=26 stringy \emph{positive}-norm vertex
operators \cite{12}, we will first tabulate Young diagrams of D=26 \textit{%
zero}-norm states at each mass level given Young diagrams of positive-norm
states at the same mass level. A consistent check of counting of number of
zero-norm states by using the background ghost fields in WSFT \ was given in 
\cite{6}. Here we go one step further and invent a simplified method to
explicitly construct D=26 stringy \emph{zero}-norm states. As examples, we
calculate all relevant zero-norm states up to the spin-five level. General%
\texttt{\ }formulas of some zero-norm tensor states at an arbitrary mass
levels will also be given. In section III, we then use these zero-norm
states and their corresponding stringy Ward identities, together with the 
\textit{factorization }property of stringy vertex operators, to explicitly
show the reduction of string-tree scattering amplitudes of degenerate
positive-norm propagating states up to the spin-five level. This calculation
justifies two previous independent calculations based on the massive
worldsheet sigma-model approach \cite{3} and WSFT approach \cite{6}.

\section{Calculation of zero-norm states}

The vertex operator of a physical state of open bosonic string

\begin{equation}
\left\vert \Psi \right\rangle =\sum C_{\mu _{1}...\mu _{m}}\alpha
_{-n_{1}}^{\mu _{1}}...\alpha _{-n_{m}}^{\mu _{m}}\left\vert
0;k\right\rangle ,[\alpha _{m}^{\mu },\alpha _{n}^{\nu }]=m\eta ^{\mu \nu
}\delta _{m+n}
\end{equation}%
is given by \cite{13}

\begin{equation}
\Psi (z)=\sum C_{\mu _{1}...\mu _{m}}N_{m}:\prod (\partial
_{z}^{n_{j}}x^{\mu _{j}})e^{ik\cdot X(z)}:,
\end{equation}%
where $N_{m}=i^{m}\prod \{(n_{j}-1)!\}^{-1}$. In the OCFQ spectrum, physical
states in eq.(1) are subject to the following Virasoro conditions

\begin{equation}
(L_{0}-1)\left\vert \Psi \right\rangle =0,L_{1}\left\vert \Psi \right\rangle
=L_{2}\left\vert \Psi \right\rangle =0,  \tag{3a,b}
\end{equation}%
where

\begin{equation}
L_{m}=\frac{1}{2}\sum_{-\infty }^{\infty }:\alpha _{m-n}\cdot \alpha _{n}: 
\tag{4}
\end{equation}%
and $\alpha _{0}\equiv k$. The solutions of eqs.(3a,b) include positive-norm
propagating states and two types of zero-norm states. The latter are \cite%
{14}

\begin{equation}
\text{Type I}:L_{-1}\left| x\right\rangle ,\text{ where }L_{1}\left|
x\right\rangle =L_{2}\left| x\right\rangle =0,\text{ }L_{0}\left|
x\right\rangle =0;  \tag{5}
\end{equation}

\begin{equation}
\text{Type II}:(L_{-2}+\frac{3}{2}L_{-1}^{2})\left\vert \widetilde{x}%
\right\rangle ,\text{ where }L_{1}\left\vert \widetilde{x}\right\rangle
=L_{2}\left\vert \widetilde{x}\right\rangle =0,\text{ }(L_{0}+1)\left\vert 
\widetilde{x}\right\rangle =0.  \tag{6}
\end{equation}%
Equations (5) and (6) can be derived from Kac determinant in conformal field
theory. While type I states have zero-norm at any spacetime dimension, type
II states have zero-norm \textit{only} at D=26. The existence of type II
zero-norm states signals the importance of zero-norm states in the structure
of the theory of string. It is straightforward to solve positive-norm state
solutions of eq.(3a, b) for some low-lying states, but soon becomes
practically unmanageable. The authors of Ref \cite{12} gave a simple
prescription to solve the positive-norm state solutions of eq.(3a, b). The
strategy is to apply the Virasoro conditions only to purely transverse
states, so that the zero-norm states will be got rid of at the very
beginning. This prescription simplified a lot of computation although some
complexities remained for low spin states at higher levels. Our aim here, on
the contrary, is to generate zero-norm states in eqs.(5) and (6), so that
all physical state solutions of eq.(3) will be completed.

Let's first assume we are given positive-norm state solutions of some mass
level $n$. The number of positive-norm degree of freedom at mass level $n$ ( 
$M^{2}=2(n-1)$) is given by $N_{24}(n)$, where \cite{15}

\begin{equation}
N_{D}(n)=\frac{1}{2\pi i}\doint \frac{dx}{x^{n+1}}(\tprod_{k=1}^{\infty }%
\frac{1}{1-x^{k}})^{D}.  \tag{7}
\end{equation}%
On the other hand, the number of physical state degree of freedom is given
by $N_{25}(n)$ in view of the constraints in eq (3a,b). The discrepancy is
of course due to physical zero-norm states given by solutions of eqs.(5) and
(6). That is, among 25 chains of $\alpha _{m}^{\mu }$ oscillators one chain
forms zero-norm states. Thus we can easily tabulate Young diagrams of
zero-norm states at each mass level given Young diagrams of positive-norm
states at the same mass level calculated by the simplified prescription in 
\cite{12}. For example, positive-norm state $\raisebox{0.06in}{\fbox{%
\rule[0.04cm]{0.04cm}{0cm}}}\raisebox{0.06in}{\fbox{%
\rule[0.04cm]{0.04cm}{0cm}}}\raisebox{0.06in}{\fbox{%
\rule[0.04cm]{0.04cm}{0cm}}}\raisebox{0.06in}{\fbox{%
\rule[0.04cm]{0.04cm}{0cm}}}$ \ at mass level $n=4$ gives zero-norm states $%
\raisebox{0.06in}{\fbox{\rule[0.04cm]{0.04cm}{0cm}}}\raisebox{0.06in}{\fbox{%
\rule[0.04cm]{0.04cm}{0cm}}}\raisebox{0.06in}{\fbox{%
\rule[0.04cm]{0.04cm}{0cm}}}$+$\raisebox{0.06in}{\fbox{%
\rule[0.04cm]{0.04cm}{0cm}}}\raisebox{0.06in}{\fbox{%
\rule[0.04cm]{0.04cm}{0cm}}}$+$\raisebox{0.06in}{\fbox{%
\rule[0.04cm]{0.04cm}{0cm}}}+\bullet $ , posive-norm state $%
\raisebox{0.06in}{\fbox{\rule[0.04cm]{0.04cm}{0cm}}}\hspace{-0.094in}\hspace{%
-0.04cm}\raisebox{-.047in}{\fbox{\rule[0.04cm]{0.04cm}{0cm}}}\hspace{-0.006in%
}\hspace{-0.006in}\hspace{0.02cm}\raisebox{0.06in}{\fbox{%
\rule[0.04cm]{0.04cm}{0cm}}}$ gives zero-norm states $\raisebox{0.06in}{%
\fbox{\rule[0.04cm]{0.04cm}{0cm}}}\hspace{-0.094in}\hspace{-0.04cm}%
\raisebox{-.047in}{\fbox{\rule[0.04cm]{0.04cm}{0cm}}}+\raisebox{0.06in}{%
\fbox{\rule[0.04cm]{0.04cm}{0cm}}}\raisebox{0.06in}{\fbox{%
\rule[0.04cm]{0.04cm}{0cm}}}+\raisebox{0.06in}{\fbox{%
\rule[0.04cm]{0.04cm}{0cm}}}$ \ and positive-norm state $\raisebox{0.06in}{%
\fbox{\rule[0.04cm]{0.04cm}{0cm}}}\raisebox{0.06in}{\fbox{%
\rule[0.04cm]{0.04cm}{0cm}}}$ gives zero-norm states $\raisebox{0.06in}{%
\fbox{\rule[0.04cm]{0.04cm}{0cm}}}+\bullet .$ This completes the zero-norm
states at mass level $n=4.$ Young diagrams of zero-norm states up to mass
level $M^{2}=10$, together with positive-norm states calculated in \cite{12}%
, are listed in the Appendix A. A consistent check of counting of zero-norm
states by using background ghost fields in WSFT \ was given in \cite{6}.

To explicitly calculate zero-norm states is another issue. Suppose we are
given some low-lying positive-norm state solutions. It is interesting to see
the similarity between eq.(3a, b) and eqs.(5) and (6) for $\left\vert
x\right\rangle $ and $\left\vert \widetilde{x}\right\rangle $. The only
difference is the \textquotedblright mass shift\textquotedblright\ of $L_{0}$
equations. As is well-known, the $L_{1\text{ }}$and $L_{2}$ equations give
the transverse and traceless conditions on the spin polarization. It turns
out that, in many cases, the $L_{1\text{ }}$and $L_{2}$ equations will not
refer to the $L_{0}$ equation or on-mass-shell condition. In these cases, a
positive-norm state solution for $\left\vert \Psi \right\rangle $ at mass
level $n$ will give a zero-norm state solution $L_{-1}\left\vert
x\right\rangle $ at mass level $n+1$ simply by taking $\left\vert
x\right\rangle =\left\vert \Psi \right\rangle $ and shifting $k^{2}$ by one
unit. Similarly, one can easily get a type II zero-norm state $(L_{-2}+\frac{%
3}{2}L_{-1}^{2})\left\vert \widetilde{x}\right\rangle $ at mass level $n+2$
simply by taking $\left\vert \widetilde{x}\right\rangle =\left\vert \Psi
\right\rangle $ and shifting $k^{2}$ by two units. For those cases where $%
L_{1\text{ }}$and $L_{2}$ equations do refer to $L_{0}$ equation, our
prescription needs to be modified. We will give some examples to illustrate
this method. Note that once we generate a zero-norm state, it soon becomes \
a candidate of physical state $\left\vert \Psi \right\rangle $ to generate
two new zero-norm states at even higher levels.

1. The \ first zero-norm state begin at $k^{2}=0$. This state is suggested
from the positive-norm tachyon state $\left\vert 0,k\right\rangle $ with $%
k^{2}$ $=2.$ Taking $\left\vert x\right\rangle =\left\vert 0,k\right\rangle $
and shifting $k^{2}$ by one unit to $k^{2}=0$, we get a type I zero-norm
state.

\begin{equation}
L_{-1\text{ }}\left\vert x\right\rangle =k\cdot \alpha _{-1}\left\vert
0,k\right\rangle ;\left\vert x\right\rangle =\left\vert 0,k\right\rangle
,-k^{2}=M^{2}=0.  \tag{8}
\end{equation}

2. At the first massive level $k^{2}=-2,$ tachyon suggests a type II
zero-norm state

\begin{equation}
(L_{-2}+\frac{3}{2}L_{-1}^{2})\left\vert \widetilde{x}\right\rangle =[\frac{1%
}{2}\alpha _{-1}\cdot \alpha _{-1}+\frac{5}{2}k\cdot \alpha _{-2}+\frac{3}{2}%
(k\cdot \alpha _{-1})^{2}]\left\vert 0,k\right\rangle ;\left\vert \widetilde{%
x}\right\rangle =\left\vert 0,k\right\rangle ,-k^{2}=2.  \tag{9}
\end{equation}%
Positive-norm massless vector state suggests a type I zero-norm state

\begin{equation}
L_{-1}\left\vert x\right\rangle =[\theta \cdot \alpha _{-2}+(k\cdot \alpha
_{-1})(\theta \cdot \alpha _{-1})]\left\vert 0,k\right\rangle ;\left\vert
x\right\rangle =\theta \cdot \alpha _{-1}\left\vert 0,k\right\rangle
,-k^{2}=2,\theta \cdot k=0.  \tag{10}
\end{equation}%
However, massless singlet zero-norm state (8) does not give a type I
zero-norm state at the first massive level $k^{2}=-2$ since $L_{1\text{ }}$
equation on state (8) refers to $L_{0}$ equation, $k^{2}=0.$ This means that 
$L_{1}$ will not annihilate state (8) if one shifts the mass to $k^{2}=-2.$

3. At the second massive level $k^{2}=-4,$ positive-norm massless vector
state suggests a type II zero-norm state

\begin{eqnarray}
(L_{-2}+\frac{3}{2}L_{-1}^{2})\left\vert \widetilde{x}\right\rangle
&=&\{4\theta \cdot \alpha _{-3}+\frac{1}{2}(\alpha _{-1}\cdot \alpha
_{-1})(\theta \cdot \alpha _{-1})+\frac{5}{2}(k\cdot \alpha _{-2})(\theta
\cdot \alpha _{-1})  \notag \\[0.01in]
&&+\frac{3}{2}(k\cdot \alpha _{-1})^{2}(\theta \cdot \alpha _{-1})+3(k\cdot
\alpha _{-1})(\theta \cdot \alpha _{-2})\}\left\vert 0,k\right\rangle ; 
\notag \\
\left\vert \widetilde{x}\right\rangle &=&\theta \cdot \alpha _{-1}\left\vert
0,k\right\rangle ,-k^{2}=4,k\cdot \theta =0.  \TCItag{11}
\end{eqnarray}%
However, massless singlet zero-norm state (8) does not give a type II
zero-norm state at mass level $k^{2}=-4$ for the same reason stated after eq
(10). Positive-norm spin-two state at $k^{2}=-2$ suggests a type I zero-norm
state

\begin{eqnarray}
L_{-1}\left\vert x\right\rangle &=&[2\theta _{\mu \nu }\alpha _{-1}^{\mu
}\alpha _{-2}^{\nu }+k_{\lambda }\theta _{\mu \nu }\alpha _{-1}^{\lambda \mu
\nu }]\left\vert 0,k\right\rangle ;\left\vert x\right\rangle =\theta _{\mu
\nu }\alpha _{-1}^{\mu \nu }\left\vert 0,k\right\rangle ,-k^{2}=4,  \notag \\
k\cdot \theta &=&\eta ^{\mu \nu }\theta _{\mu \nu }=0,\theta _{\mu \nu
}=\theta _{\nu \mu },  \TCItag{12}
\end{eqnarray}
where $\alpha _{-1}^{\lambda \mu \nu }\equiv \alpha _{-1}^{\lambda }\alpha
_{-1}^{\mu }\alpha _{-1}^{\nu }.$ Similar notations will be used in the rest
of this paper. Vector zero-norm state with $k^{2}=-2$ in eq.(10) does not
give a type I zero-norm state for the same reason stated after eq.(10). In
this case, however, one can modify $\left\vert x\right\rangle $ to be

\begin{equation}
\text{Ansatz: }\left\vert x\right\rangle =[a\theta \cdot \alpha
_{-2}+b(k\cdot \alpha _{-1})(\theta \cdot \alpha _{-1})]\left\vert
0,k\right\rangle ;-k^{2}=4,\theta \cdot k=0,  \tag{13}
\end{equation}%
where $a,b$ are undetermined constants. $L_{0}$ equation is then trivially
satisfied and $L_{1},$ $L_{2}$ equations give $a:b=2:1$. This gives a type I
zero-norm state

\begin{eqnarray}
L_{-1}\left\vert x\right\rangle &=&[\frac{1}{2}(k\cdot \alpha
_{-1})^{2}(\theta \cdot \alpha _{-1})+2\theta \cdot \alpha _{-3}+\frac{3}{2}%
(k\cdot \alpha _{-1})(\theta \cdot \alpha _{-2})  \notag \\
&&+\frac{1}{2}(k\cdot \alpha _{-2})(\theta \cdot \alpha _{-1})]\left\vert
0,k\right\rangle ;-k^{2}=4,\theta \cdot k=0.  \TCItag{14}
\end{eqnarray}%
Similarly, we modify the singlet zero-norm state with $k^{2}=-2$ in eq.(9)
to be

\begin{equation}
\text{Ansatz: }\left\vert x\right\rangle =[\frac{5}{2}ak\cdot \alpha _{-2}+%
\frac{1}{2}b\alpha _{-1}\cdot \alpha _{-1}+\frac{3}{2}c(k\cdot \alpha
_{-1})^{2}]\left\vert 0,k\right\rangle ;-k^{2}=4,  \tag{15}
\end{equation}%
where $a$, $b$ and $c$ are undetermined constants. $L_{1}$ and $L_{2}$
equations give

\begin{equation}
5a+b+3k^{2}c=0,5k^{2}a+13b+\frac{3}{2}k^{2}c=0.  \tag{16}
\end{equation}%
For $k^{2}=-4$, we \ have $a:b:c=5:9:\frac{17}{6}$. This gives a type I
zero-norm state

\begin{eqnarray}
L_{-1}\left\vert x\right\rangle &=&[\frac{17}{4}(k\cdot \alpha _{-1})^{3}+%
\frac{9}{2}(k\cdot \alpha _{-1})(\alpha _{-1}\cdot \alpha _{-1})+9(\alpha
_{-1}\cdot \alpha _{-2})  \notag \\
&&+21(k\cdot \alpha _{-1})(k\cdot \alpha _{-2})+25(k\cdot \alpha
_{-3})]\left\vert 0,k\right\rangle ;  \TCItag{17} \\
-k^{2} &=&4.  \notag
\end{eqnarray}%
This completes the four zero-norm states at the second massive level. Note
that state (17) was calculated in Ref \cite{7} without modification. The
coefficients there thus need to be corrected although the main results
remain valid. It is interesting to note that the Young tableau of zero-norm
states at level $M^{2}=4$ are the sum of those of all physical states at two
lower levels, $M^{2}=2$ and $M^{2}=0$, \textit{except }the singlet zero-norm
state due to the dependence of $L_{1}$ and $L_{2}$ equations on$\ L_{0}$
condition in state (8). For those cases that $L_{1}$\ and $L_{2}$\ equations
not referring to $L_{0}$\ condition, our construction gives us a very simple
way to calculate zero-norm states at any mass level \ $n$\ given those of
positive-norm states at lower levels constructed by the simplified method in
Ref \cite{12}. When the modified method was needed to calculate a higher
mass level zero-norm state from a lower mass level physical state like
eq.(8), an inconsistency may result and one gets no zero-norm state. This
explains the discrepancy of singlet zero-norm states at levels $M^{2}=2,4$,$%
8 $ and a vector zero-norm state at level $M^{2}=10$.

4. Similar method can be used to calculate zero-norm states at level $%
M^{2}=6.$ We will just list those which are relevant for the discussion in
section III. They are (from now on, unless otherwise stated, each spin
polarization is assumed to be transverse, traceless and is symmetric with
respect to each group of indices as in Ref \cite{12})

\begin{equation}
L_{-1}\left\vert x\right\rangle =\theta _{\mu \nu \lambda }(k_{\beta }\alpha
_{-1}^{\mu \nu \lambda \beta }+3\alpha _{-1}^{\mu \nu }\alpha _{-2}^{\lambda
})\left\vert 0,k\right\rangle ;\left\vert x\right\rangle =\theta _{\mu \nu
\lambda }\alpha _{-1}^{\mu \nu \lambda }\left\vert 0,k\right\rangle , 
\tag{18}
\end{equation}

\begin{equation}
L_{-1}\left\vert x\right\rangle =[k_{\lambda }\theta _{\mu \nu }\alpha
_{-1}^{\mu _{\lambda }}\alpha _{-2}^{\nu }+2\theta _{\mu \nu }\alpha
_{-1}^{\mu }\alpha _{-3}^{\nu }\left\vert 0,k\right\rangle ;\left\vert
x\right\rangle =\theta _{\mu \nu }\alpha _{-1}^{\mu }\alpha _{-2}^{\nu
}\left\vert 0,k\right\rangle ,\text{ where }\theta _{\mu \nu }=-\theta _{\nu
\mu },  \tag{19}
\end{equation}

\begin{eqnarray}
L_{-1}\left\vert x\right\rangle &=&[2\theta _{\mu \nu }\alpha _{-2}^{\mu \nu
}+4\theta _{\mu \nu }\alpha _{-1}^{\mu }\alpha _{-3}^{\nu }+2(k_{\lambda
}\theta _{\mu \nu }+k_{(\lambda }\theta _{\mu \nu )})\alpha _{-1}^{\lambda
\mu }\alpha _{-2}^{\nu }+\frac{2}{3}k_{\lambda }k_{\beta }\theta _{\mu \nu
}\alpha _{-1}^{\mu \nu \lambda \beta }]\left\vert 0,k\right\rangle ;  \notag
\\
\left\vert x\right\rangle &=&[2\theta _{\mu \nu }\alpha _{-1}^{\mu }\alpha
_{-2}^{\nu }+\frac{2}{3}k_{\lambda }\theta _{\mu \nu }\alpha _{-1}^{\mu \nu
\lambda }]\left\vert 0,k\right\rangle  \TCItag{20}
\end{eqnarray}%
and

\begin{eqnarray}
(L_{-2}+\frac{3}{2}L_{-1}^{2})\left\vert \widetilde{x}\right\rangle
&=&[3\theta _{\mu \nu }\alpha _{-2}^{\mu \nu }+8\theta _{\mu \nu }\alpha
_{-1}^{\mu }\alpha _{-3}^{\nu }+(k_{\lambda }\theta _{\mu \nu }+\frac{15}{2}%
k_{(\lambda }\theta _{\mu \nu )})\alpha _{-1}^{\lambda \mu }\alpha
_{-2}^{\nu }  \notag \\
&&+(\frac{1}{2}\eta _{\lambda \beta }\theta _{\mu \nu }+\frac{3}{2}%
k_{\lambda }k_{\beta }\theta _{\mu \nu })\alpha _{-1}^{\mu \nu \lambda \beta
}]\left\vert 0,k\right\rangle ;  \notag \\
\left\vert \widetilde{x}\right\rangle &=&\theta _{\mu \nu }\alpha _{-1}^{\mu
\nu }\left\vert 0,k\right\rangle .  \TCItag{21}
\end{eqnarray}%
Note that $\left\vert x\right\rangle $ in eq.(20) has been modified as we
did for eq (13). To further illustrate our method, we calculate the type I
singlet zero-norm state from eq.(17) as following

\begin{eqnarray}
\text{Ansatz} &:&\left\vert x\right\rangle =[a(k\cdot \alpha
_{-1})^{3}+b(k\cdot \alpha _{-1})(\alpha _{-1}\cdot \alpha _{-1})+c(k\cdot
\alpha _{-1})(k\cdot \alpha _{-2})  \notag \\
&&+d(\alpha _{-1}\cdot \alpha _{-2})+f(k\cdot \alpha _{-3})\left\vert
0,k\right\rangle ;  \notag \\
-k^{2} &=&6.  \TCItag{22}
\end{eqnarray}%
The $L_{1}$ and $L_{2}$ equations can be easily used to determine $%
a:b:c:d:f=37:72:261:216:450.$ This gives the type I singlet zero-norm state

\begin{eqnarray}
L_{-1}\left\vert x\right\rangle &=&[a(k\cdot \alpha _{-1})^{4}+b(k\cdot
\alpha _{-1})^{2}(\alpha _{-1}\cdot \alpha _{-1})+(2b+d)(k\cdot \alpha
_{-1})(\alpha _{-1}\cdot \alpha _{-2})  \notag \\
&&+(c+3a)(k\cdot \alpha _{-1})^{2}(k\cdot \alpha _{-2})+c(k\cdot \alpha
_{-2})^{2}+d(\alpha _{-2}\cdot \alpha _{-2})+b(k\cdot \alpha _{-2})(\alpha
_{-1}\cdot \alpha _{-1})  \notag \\
&&+(2c+f)(k\cdot \alpha _{-3})(k\cdot \alpha _{-1})+2d(\alpha _{-1}\cdot
\alpha _{-3})+3f(k\cdot \alpha _{-4})]\left\vert 0,k\right\rangle ,  \notag
\\
-k^{2} &=&6.  \TCItag{23}
\end{eqnarray}

5. We list relevant zero-norm states at level $M^{2}=8$ from the known
positive-norm states and zero-norm states at level $M^{2}=4,6.$ They are

\begin{equation}
L_{-1}\left\vert x\right\rangle =(k_{\beta }\theta _{\mu \nu \lambda \gamma
}\alpha _{-1}^{\mu \nu \lambda \gamma \beta }+4\theta _{\mu \nu \lambda
\gamma }\alpha _{-1}^{\mu \nu \lambda }\alpha _{-2}^{\gamma }\left\vert
0,k\right\rangle ;\left\vert x\right\rangle =\theta _{\mu \nu \lambda \gamma
}\alpha _{-1}^{\mu \nu \lambda \gamma }\left\vert 0,k\right\rangle , 
\tag{24}
\end{equation}

\begin{eqnarray}
L_{-1}\left\vert x\right\rangle &=&\theta _{\mu \nu \lambda }[\frac{3}{4}%
k_{\beta }k_{\gamma }\alpha _{-1}^{\mu \nu \lambda \gamma \beta }+3k_{\beta
}\alpha _{-1}^{\mu \nu \beta }\alpha _{-2}^{\lambda }+3k_{\beta }\alpha
_{-1}^{(\mu \nu \lambda }\alpha _{-2}^{\beta )}+6\alpha _{-1}^{(\mu }\alpha
_{-2}^{\nu \lambda )}  \notag \\
+6\alpha _{-1}^{(\mu \nu }\alpha _{-3}^{\lambda )}]\left\vert
0,k\right\rangle ;\text{ }\left\vert x\right\rangle &=&\theta _{\mu \nu
\lambda }(\frac{3}{4}k_{\beta }\alpha _{-1}^{\mu \nu \lambda \beta }+3\alpha
_{-1}^{\mu \nu }\alpha _{-2}^{\lambda })\left\vert 0,k\right\rangle , 
\TCItag{25}
\end{eqnarray}

\begin{eqnarray}
(L_{-2}+\frac{3}{2}L_{-1}^{2})\left\vert \widetilde{x}\right\rangle
&=&\theta _{\mu \nu \lambda }[(\frac{3}{2}k_{\beta }k_{\gamma }+\frac{1}{2}%
\eta _{\gamma \beta })\alpha _{-1}^{\mu \nu \lambda \beta \gamma }+k_{\gamma
}(\frac{1}{2}\alpha _{-1}^{\mu \nu \lambda }\alpha _{-2}^{\gamma }+8\alpha
_{-1}^{(\mu \nu \lambda }\alpha _{-2}^{\gamma )})  \notag \\
&&+3\alpha _{-1}^{(\mu }\alpha _{-2}^{\nu \lambda )}+6\alpha _{-1}^{(\mu \nu
}\alpha _{-3}^{\lambda )}]\left\vert 0,k\right\rangle ;  \notag \\
\left\vert \widetilde{x}\right\rangle &=&\theta _{\mu \nu \lambda }\alpha
_{-1}^{\mu \nu \lambda }\left\vert 0,k\right\rangle ,  \TCItag{26}
\end{eqnarray}

\begin{eqnarray}
L_{-1}\left\vert x\right\rangle &=&\theta _{\mu \nu ,\lambda }(k_{\gamma
}\alpha _{-1}^{\gamma \mu \nu }\alpha _{-2}^{\lambda }+2\alpha _{-1}^{\mu
}\alpha _{-2}^{\nu \lambda }+2\alpha _{-1}^{\mu \nu }\alpha _{-3}^{\lambda
})\left\vert 0,k\right\rangle ;  \notag \\
\left\vert x\right\rangle &=&\theta _{\mu \nu ,\lambda }\alpha _{-1}^{\mu
\nu }\alpha _{-2}^{\lambda }\left\vert 0,k\right\rangle ,\text{ where }%
\theta _{\mu \nu ,\lambda }\text{ is mixed symmetric,}  \TCItag{27}
\end{eqnarray}

\begin{eqnarray}
L_{-1}\left\vert x\right\rangle &=&\theta _{\mu \nu }(\frac{3}{4}k_{\beta
}k_{\lambda }\alpha _{-1}^{\beta \lambda \mu }\alpha _{-2}^{\nu
}+4k_{\lambda }\alpha _{-1}^{\lambda \mu }\alpha _{-3}^{\nu }+\frac{3}{4}%
k_{\lambda }\alpha _{-1}^{\mu }\alpha _{-2}^{\nu \lambda }+2\alpha
_{-2}^{\mu }\alpha _{-3}^{\nu }+6\alpha _{-1}^{\mu }\alpha _{-4}^{\nu
})\left\vert 0,k\right\rangle ;  \notag \\
\left\vert x\right\rangle &=&(\frac{3}{4}k_{\lambda }\alpha _{-1}^{\lambda
\mu }\alpha _{-2}^{\nu }+2\alpha _{-1}^{\mu }\alpha _{-3}^{\nu })\left\vert
0,k\right\rangle ,\text{ where }\theta _{\mu \nu }=-\theta _{\nu \mu }, 
\TCItag{28}
\end{eqnarray}%
and

\begin{eqnarray}
(L_{-2}+\frac{3}{2}L_{-1}^{2})\left\vert \widetilde{x}\right\rangle
&=&\theta _{\mu \nu }[(\frac{3}{2}k_{\gamma }k_{\lambda }+\frac{1}{2}\eta
_{\gamma \lambda })\alpha _{-1}^{\gamma \lambda \mu }\alpha _{-2}^{\nu
}+6k_{\lambda }\alpha _{-1}^{\lambda \mu }\alpha _{-3}^{\nu }+\frac{5}{2}%
k_{\lambda }\alpha _{-1}^{\mu }\alpha _{-2}^{\nu \lambda }  \notag \\
+2\alpha _{-2}^{\mu }\alpha _{-3}^{\nu }+\alpha _{-1}^{\mu }\alpha
_{-4}^{\nu }]\left\vert 0,k\right\rangle ,\left\vert \widetilde{x}%
\right\rangle &=&\theta _{\mu \nu }\alpha _{-1}^{\mu }\alpha _{-2}^{\nu
}\left\vert 0,k\right\rangle ,\text{ where }\theta _{\mu \nu }=-\theta _{\nu
\mu }.  \TCItag{29}
\end{eqnarray}%
Note that the modified method was used in eqs.(25) and (28).

6. Finally, we calculate general formulas of some zero-norm tensor states at
arbitrary mass levels by making use of general formulas of some
positive-norm states listed in Ref \cite{12}.

a. 
\begin{equation}
L_{-1}\theta _{\mu _{1}...\mu _{m}}\alpha _{-1}^{\mu _{1}...\mu
_{m}}\left\vert 0,k\right\rangle =\theta _{\mu _{1}...\mu _{m}}(k_{\lambda
}\alpha _{-1}^{\lambda \mu _{1}...\mu _{m}}+m\alpha _{-2}^{\mu _{1}}\alpha
_{-1}^{\mu _{2}...\mu _{m}})\left\vert 0,k\right\rangle ,  \tag{30}
\end{equation}
where $-k^{2}=M^{2}=2m,m=0,1,2,3...$. For example, $m=0,1$ give eqs (8) and
(10).

b. 
\begin{eqnarray}
&&(L_{-2}+\frac{3}{2}L_{-1}^{2})\theta _{\mu _{1}...\mu _{m}}\alpha
_{-1}^{\mu _{1}...\mu _{m}}\left\vert 0,k\right\rangle  \notag \\
&=&\{\theta _{\mu _{1}...\mu _{m}}[(\frac{3}{2}k_{\nu }k_{\lambda }+\frac{1}{%
2}\eta _{\nu \lambda })\alpha _{-1}^{\nu \lambda \mu _{1}...\mu _{m}}+\frac{3%
}{2}m(m-1)\alpha _{-2}^{\mu _{1}\mu _{2}}\alpha _{-1}^{\mu _{3}...\mu _{m}} 
\notag \\
&&+(1+3m)\alpha _{-1}^{\mu _{1}...\mu _{m-1}}\alpha _{-3}^{\mu _{m}}]+[\frac{%
3}{2}(m+1)k_{(\lambda }\theta _{\mu _{1}...\mu _{m})}+\frac{3}{2}mk_{\mu
_{m}}\theta _{\mu _{1}...\mu _{m-1\lambda })}]  \notag \\
&&\alpha _{-1}^{\mu _{1}...\mu _{m}}\alpha _{-2}^{\lambda }\}\left\vert
0,k\right\rangle ,  \TCItag{31}
\end{eqnarray}%
where $-k^{2}=M^{2}=2m+2,m=0,1,2...$. For example, $m=0,1$ give eqs.(9) and
(11).

c. 
\begin{eqnarray}
&&L_{-1}\theta _{\mu _{1}...\mu _{m-2},\mu _{m-1}}\alpha _{-1}^{\mu
_{1}...\mu _{m-2}}\alpha _{-2}^{\mu _{m-1}}\left\vert 0,k\right\rangle 
\notag \\
&=&\theta _{\mu _{1}...\mu _{m-2},\mu _{m-1}}[k_{\lambda }\alpha
_{-1}^{\lambda \mu _{1}...\mu _{m-2}}\alpha _{-2}^{\mu _{m-1}}+(m-2)\alpha
_{-1}^{\mu _{1}...\mu _{m-3}}\alpha _{-2}^{\mu _{m-2}\mu _{m}}  \notag \\
&&+2\alpha _{-1}^{\mu _{1}...\mu _{m-2}}\alpha _{-2}^{\mu _{m-1}}]\left\vert
0,k\right\rangle ,\hspace{0.1cm}\raisebox{0.06in}{\fbox{%
\rule[0.04cm]{0.04cm}{0cm}}}\hspace{-0.094in}\hspace{-0.04cm}%
\raisebox{-.047in}{\fbox{\rule[0.04cm]{0.04cm}{0cm}}}\hspace{-0.006in}%
\hspace{-0.006in}\hspace{0.02cm}\raisebox{0.06in}{\fbox{......%
\rule[0.01cm]{0.18cm}{0cm}}}\hspace{-0.025cm}\raisebox{0.06in}{\fbox{%
\rule[0.04cm]{0.04cm}{0cm}}}  \TCItag{32}
\end{eqnarray}%
where $-k^{2}=M^{2}=2m,m=3,4,5...$. For example, $m=3,4$ give eqs.(19) and
(27).

d. 
\begin{eqnarray}
&&(L_{-2}+\frac{3}{2}L_{-1}^{2})\theta _{\mu _{1}...\mu _{m-2},\mu
_{m-1}}\alpha _{-1}^{\mu _{1}...\mu _{m-2}}\alpha _{-2}^{\mu
_{m-1}}\left\vert 0,k\right\rangle  \notag \\
&=&\theta _{\mu _{1}...\mu _{m-2},\mu _{m-1}}[(\frac{3}{2}k_{\lambda }k_{\nu
}+\frac{1}{2}\eta _{\lambda \nu })\alpha _{-1}^{\mu _{1}...\mu _{m-2}\lambda
\nu }\alpha _{-2}^{\mu _{m-1}}+6k_{\lambda }\alpha _{-1}^{\mu _{1}...\mu
_{m-2}\lambda }\alpha _{-3}^{\mu _{m-1}}  \notag \\
&&+(\frac{3}{2}m-2)k_{\lambda }\alpha _{-1}^{\mu _{1}...\mu _{m-2}}\alpha
_{-2}^{\mu _{m-1\lambda }}+2(m-2)\alpha _{-1}^{\mu _{1}...\mu _{m-3}}\alpha
_{-2}^{\mu _{m-2}}\alpha _{-3}^{\mu _{m-1}}+11\alpha _{-1}^{\mu _{1}...\mu
_{m-2}}\alpha _{-4}^{\mu _{m-1}}  \notag \\
&&+k_{\lambda }\alpha _{-1}^{\mu _{1}...\mu _{m-3}\lambda }\alpha _{-2}^{\mu
_{m-2}\mu _{m-1}}+(m-3)\alpha _{-1}^{\mu _{1}...\mu _{m-4}}\alpha _{-2}^{\mu
_{m-3}\mu _{m-2}\mu _{m-1}}]\left\vert 0,k\right\rangle ,\hspace{0.1cm}%
\raisebox{0.06in}{\fbox{\rule[0.04cm]{0.04cm}{0cm}}}\hspace{-0.094in}\hspace{%
-0.04cm}\raisebox{-.047in}{\fbox{\rule[0.04cm]{0.04cm}{0cm}}}\hspace{-0.006in%
}\hspace{-0.006in}\hspace{0.02cm}\raisebox{0.06in}{\fbox{......%
\rule[0.01cm]{0.18cm}{0cm}}}\hspace{-0.025cm}\raisebox{0.06in}{\fbox{%
\rule[0.04cm]{0.04cm}{0cm}}}  \TCItag{33}
\end{eqnarray}
where $-k^{2}=M^{2}=2m+2,m=3,4,5...$. For example, $m=3$ gives eq.(29).

e. 
\begin{eqnarray}
&&L_{-1}\theta _{\mu _{1}...\mu _{m-4},\mu _{m-3}\mu _{m-2}}(\alpha
_{-1}^{\mu _{1}...\mu _{m-4}}\alpha _{-2}^{\mu _{m-3}\mu _{m-2}}-\frac{4}{3}%
\alpha _{-1}^{\mu _{1}...\mu _{m-3}}\alpha _{-3}^{\mu _{m-2}})  \notag \\
&=&\theta _{\mu _{1}...\mu _{m-4},\mu _{m-3}\mu _{m-2}}[k_{\lambda }\alpha
_{-1}^{\lambda \mu _{1}...\mu _{m-4}}\alpha _{-2}^{\mu _{m-3}\mu
_{m-2}}+(m-4)\alpha _{-1}^{\mu _{1}...\mu _{m-3}}\alpha _{-2}^{\mu _{m-4}\mu
_{m-3}\mu _{m-2}}  \notag \\
&&+\frac{16}{3}\alpha _{-1}^{\mu _{1}...\mu _{m-4}}\alpha _{-3}^{\mu
_{m-3}}\alpha _{-2}^{\mu _{m-2}}+\frac{4}{3}k_{\lambda }\alpha
_{-1}^{\lambda \mu _{1}...\mu _{m-3}}\alpha _{-3}^{\mu _{m-2}}+4\alpha
_{-1}^{\mu _{1}...\mu _{m-3}}\alpha _{-4}^{\mu _{m-4}}],\hspace{0.1cm}%
\raisebox{0.06in}{\fbox{\rule[0.04cm]{0.04cm}{0cm}}}\hspace{-0.094in}\hspace{%
-0.04cm}\raisebox{-.047in}{\fbox{\rule[0.04cm]{0.04cm}{0cm}}}\hspace{-0.006in%
}\hspace{-0.006in}\hspace{0.02cm}\raisebox{0.06in}{\fbox{......%
\rule[0.01cm]{0.18cm}{0cm}}}\hspace{-0.025cm}\raisebox{0.06in}{\fbox{%
\rule[0.04cm]{0.04cm}{0cm}}}  \TCItag{34}
\end{eqnarray}%
where $-k^{2}=M^{2}=2m,m=5,6...$.

f. The zero-norm states of eq.(30) can be used to generate new type I
zero-norm states by the modified method as following\bigskip 
\begin{eqnarray}
&&L_{-1}\theta _{\mu _{1}...\mu _{m}}(\frac{m}{m+1}k_{\lambda }\alpha
_{-1}^{\lambda \mu _{1}...\mu _{m}}+\alpha _{-2}^{\mu _{1}}\alpha _{-1}^{\mu
_{2}...\mu _{m}})\left\vert 0,k\right\rangle  \notag \\
&=&[\frac{m}{m+1}k_{\nu }k_{\lambda }\theta _{\mu _{1}...\mu _{m}}\alpha
_{-1}^{\nu \lambda \mu _{1}...\mu _{m}}+m(k_{(\lambda }\theta _{\mu
_{1}...\mu _{m)}}+k_{\lambda }\theta _{\mu _{1}...\mu _{m}})\alpha
_{-2}^{\mu _{1}}\alpha _{-1}^{\lambda \mu _{2}...\mu _{m}}  \notag \\
&&+m(m-1)\theta _{\mu _{1}...\mu _{m}}\alpha _{-2}^{\mu _{1}\mu _{2}}\alpha
_{-1}^{\mu _{3}...\mu _{m}}+2m\theta _{\mu _{1}...\mu _{m}}\alpha _{-3}^{\mu
_{1}}\alpha _{-1}^{\mu _{2}...\mu _{m}}]\left\vert 0,k\right\rangle , 
\TCItag{35}
\end{eqnarray}%
where $-k^{2}=M^{2}=2m+2,m=1,2,3...$. For example, $m=1,2$ and $3$ give
eqs.(14), (20) and (25). Note that the coefficient of the first term in
eq.(35) has been modified to $\frac{m}{m+1}.$ Similarly, new type II
zero-norm states can also be constructed.

These are examples of some higher spin zero-norm states at arbitrary mass
levels. As in the case of positive-norm states, the complexity of the
calculation increases when calculating lower spin zero-norm states for
higher levels. Fortunately, for our purpose in this paper, it is usually
good enough to calculate higher spin zero-norm states as it will become
clear in the next section. For those formulas with transverse trace \cite{12}

\begin{equation}
\eta _{\mu \nu }^{T}=\eta _{\mu \nu }-k_{\mu }k_{\nu }/k^{2},  \tag{36}
\end{equation}%
the modified method should be used, and we have no general formulas for them.

Each zero-norm state calculated in this section corresponds to an on-shell
Ward identity, which can be easily written down. As an interesting example 
\cite{8} to illustrate the importance of zero-norm state, the \textit{%
inter-particle Ward identity} for two propagating states at the second
massive level ( $M^{2}=4$) was calculated to be ( $k\cdot \theta =0)$

\begin{equation}
(\frac{1}{2}k_{\mu }k_{\nu }\theta _{\lambda }+2\eta _{\mu \nu }\theta
_{\lambda })\mathcal{T}_{2,\chi }^{(\mu \nu \lambda )}+9k_{\mu }\theta _{\nu
}\mathcal{T}_{2,\chi }^{[\mu \nu ]}-6\theta _{\mu }\mathcal{T}_{2,\chi
}^{\mu }=0,  \tag{37}
\end{equation}%
where we have chosen , say, $v_{1}(k_{1})$\ to be the vertex operator
constructed from $D_{2}$ vector zero-norm state obtained by antisymmetrizing
those terms which contain $\alpha _{-1}^{\mu }\alpha _{-2}^{\nu }$ in the
original type I, eq.(14), and type II, eq.(11), vector zero-norm states and $%
k_{\mu }\equiv k_{1\mu }$ . Note that $v_{2},v_{3}$ and $v_{4}$ can be any
string states (including zero-norm states), and we have omitted their tensor
index for the cases of excited string states in eq (37). $\mathcal{T}%
_{2,\chi }^{\prime }s$ in eq(37) are the second massive level, $\chi $-th
order string-loop amplitudes. At this point, \{$\mathcal{T}_{2,\chi }^{(\mu
\nu \lambda )},\mathcal{T}_{2,\chi }^{(\mu \nu )},\mathcal{T}_{2,\chi }^{\mu
}$\} is identified to be the \emph{amplitude triplet} of the spin-three
state and $T^{[\mu \nu ]}$ is identified to be the amplitude of the
antisymmetric spin-two state \cite{8}. Eq.(37) thus relates the scattering
amplitudes of two different string states at the second massive level. It is
important to note that eq.(37) is, in contrast to the high-energy $\alpha
^{\prime }\rightarrow \infty $ result of Gross, valid to all string-loop and 
\emph{all} energy $\alpha ^{\prime }$, and its coefficients do depend on the
center of mass scattering angle $\phi _{CM}$, which is defined to be the
angle between $\overrightarrow{k}_{1}$ and $\overrightarrow{k}_{3}$, through
the dependence of momentum $k$ . This angular dependence disappears in the
high-energy limit of eq.(37) \cite{9}, which is consistent with Gross's
result. The inter-particle gauge symmetry corresponding to eq.(37) can be
calculated to be \cite{7}

\begin{equation}
\delta C_{(\mu \nu \lambda )}=(\frac{1}{2}\partial _{(\mu }\partial _{\nu
}\theta _{\lambda )}-2\eta _{(\mu \nu }\theta _{\lambda )}),\delta C_{[\mu
\nu ]}=9\partial _{\lbrack \mu }\theta _{\nu ]},  \tag{38}
\end{equation}%
where $\partial _{\nu }\theta ^{\nu }=0,(\partial ^{2}-4)\theta ^{\nu }=0$
are the on-shell conditions of the $D_{2}$ vector zero-norm state. $C_{(\mu
\nu \lambda )}$ and $C_{[\mu \nu ]}$ are the background fields of the
symmetric spin-three and antisymmetric spin-two states respectively at the
second mass level. Eq.(38) is the result of the first order weak field
approximation but valid to \emph{all} energy $\alpha ^{\prime }$ in the
generalized $\sigma $-model approach. It is important to note that the
decoupling of $D_{2}$ vector zero-norm state implies simultaneous change of
both $C_{(\mu \nu \lambda )}$ and $C_{[\mu \nu ]}$ , thus they form a gauge
multiplet. This important stringy phenomenon can also be justified in WSFT 
\cite{6,8}. A second order weak field calculation implies an even more
interestng spontaneously broken inter-mass level symmetry in string theory 
\cite{16}. \ 

\section{Reduction of degenerate state's amplitude}

The decoupling of degenerate positive-norm states was first discovered in
Ref \cite{3} by using generalized sigma-model approach. It was recently
justified by using WSFT for the open string case up to the spin-five level 
\cite{6}. This stringy phenomenon begins to show up at spin-four level of
open bosonic string. The explicit form of four positive-norm states at
spin-four level can be found in Ref \cite{13}. According to the decoupling
conjecture, the spin-two and the scalar positive-norm states should be
decoupled. That is, their amplitudes are determined from those of two other
higher spin states. Let's begin the discussion by first making an important
observation. According to eq.(2), the vertex operator corresponding to $%
\alpha _{-1}^{\mu \nu \lambda \gamma }$ is $A^{\mu \nu \lambda \gamma }=$ :$%
\partial x^{\mu }\partial x^{\nu }\partial x^{\lambda }\partial x^{\gamma
}e^{ik\cdot x}:$. \textit{Due to the factorization structure of this tensor
vertex, which results from the strong constraint of 2D worldsheet conformal
symmetry, the amplitude corresponding to }$A^{\mu \nu \lambda \gamma }$%
\textit{\ is fixed by its traceless, transverse spin part }$\epsilon _{\mu
\nu \lambda \gamma }$. In particular, the longitudinal parts of \ $A^{\mu
\nu \lambda \gamma }$ are determined by \textit{\ }$\epsilon _{\mu \nu
\lambda \gamma }$ through the Lorentz extension; and the trace parts of $%
A^{\mu \nu \lambda \gamma }$ are fixed by the \textit{conformal extension}.
This means that given the on-shell amplitude of $\epsilon _{\mu \nu \lambda
\gamma },$ the amplitude $\mathcal{T}_{\mu \nu \lambda \gamma }$ of $\
A_{\mu \nu \lambda \gamma }$ is fixed. Here $\mathcal{T}^{\mu \nu \lambda
\gamma }$ is defined to be the four-point function containing the \textit{%
rank}-four tensor : $\partial x^{\mu }\partial x^{\nu }\partial x^{\lambda
}\partial x^{\gamma }e^{ik\cdot x}:$ and three tachyons. Let's use a simpler
rank-two tensor to illustrate the \textit{trace fixing} or conformal
extension. Given a \textit{factorized} symmetric rank-two tensor constructed
from a $D$-vector $a^{\mu }$

\begin{eqnarray}
A^{\mu \nu } &=&a^{\mu }a^{\nu }+c\eta ^{\mu \nu }  \notag \\
&=&(a^{\mu }a^{\nu }-\frac{a^{2}}{D}\eta ^{\mu \nu })+(\frac{a^{2}}{D}%
+c)\eta ^{\mu \nu },  \TCItag{39}
\end{eqnarray}%
where we have decomposed $A^{\mu \nu }$ into a traceless spin part and a
trace part containing a scalar $c$ independent of the spin part, the trace
part of $A^{\mu \nu }$ is not fixed by the spin part of $A^{\mu \nu }.$ Now
for the \textit{homogeneous} factorized tensor, $c=0$ in eq.(39). The
traceless spin part of eq.(39) gives us $\frac{D(D+1)}{2}$ components which
is of order $D^{2}$, while the \textit{factorized} symmetric rank-two tensor 
$A^{\mu \nu }$ contains only $D$ independent components which are components
of $a^{\mu }$. It is thus easy to see that the trace part of $A^{\mu \nu }$
is fixed by the spin part of the tensor. Thus, knowing the spin part of $%
A^{\mu \nu }$ means knowing the whole tensor. This result can be easily
generalized to the decomposition of a homogeneous factorized tensor $A^{\mu
\nu }=a^{\mu }b^{\nu }$, which contains only $2D$ independent components in
contrast to the number of components of the spin part ,which is of the order 
$D^{2}.$ Similar results can be obtained for \textit{homogeneous factorized}
higher rank tensors. Note that this factorized property can only be seen in
the first order weak field approximation \cite{3} (or vertex operator
consideration), and does not show up in the zeroth order spectrum.

With the observation discussed above in mind, we can now discuss the
decoupling phenomenon at level four. It was pointed out \cite{3} that the
positive-norm spin-two state can be gauged to a gauge which contains only $%
\alpha _{-1}^{\mu \nu \lambda \gamma }$ and $\alpha _{-1}^{\mu \nu }\alpha
_{-2}^{\lambda }$ terms by making use of the gauge transformations induced
by the type I and the type II spin-two zero-norm states, eqs.(20) and (21),
to be

\begin{equation}
\lbrack (\frac{1}{3}k_{\lambda }\epsilon _{\mu \nu }+\frac{1}{2}k_{(\lambda
}\epsilon _{\mu \nu )})\alpha _{-1}^{\lambda \mu }\alpha _{-2}^{\nu }+(\frac{%
13}{174}k_{\alpha }k_{\beta }\epsilon _{\mu \nu }+\frac{3}{58}\eta _{\alpha
\beta }\epsilon _{\mu \nu })\alpha _{-1}^{\mu \nu \alpha \beta }]\left\vert
0,k\right\rangle ,  \tag{40}
\end{equation}%
where $\epsilon _{\mu \nu }$ is a symmetric traceless and transverse
spin-two tensor. Since the rank-four amplitude $\mathcal{T}_{3,\chi }^{\mu
\nu \alpha \beta }$ is fixed by the spin-four amplitude and the
mixed-symmetric rank-three amplitude $\mathcal{T}_{3,\chi }^{\lambda \mu \nu
}$ is fixed by the mixed-symmetric spin-three amplitude, the amplitude of
the spin-two state in eq.(40) is determined by those of the spin-four and
the mixed-symmetric spin-three states. (Note that $\mathcal{T}_{3,\chi
}^{(\lambda \mu \nu )}$ is fixed by the spin-four amplitude $\mathcal{T}%
_{3,\chi }^{\mu \nu \alpha \beta }$ due to the existence of a totally
symmetric spin-three zero-norm state eq.(18) at this level.) In fact, $%
\mathcal{T}_{3,\chi }^{\mu \nu \alpha \beta }$ with $\chi =1$ can be
explicitly calculated to be \cite{8}

\begin{eqnarray}
\mathcal{T}_{3,1}^{\mu \nu \lambda \gamma } &=&\frac{\Gamma (-\frac{s}{2}%
-1)\Gamma (-\frac{t}{2}-1)}{\Gamma (\frac{u}{2}+2)}[(\frac{s^{2}}{4}-s)(%
\frac{s^{2}}{4}-1)k_{3}^{\mu }k_{3}^{\nu }k_{3}^{\lambda }k_{3}^{\gamma }-t(%
\frac{t^{2}}{4}-1)(s+2)k_{1}^{(\mu }k_{1}^{\nu }k_{1}^{\lambda
}k_{3}^{\gamma )}  \notag \\
&&+\frac{3st}{2}(\frac{s}{2}+1)(\frac{t}{2}+1)k_{1}^{(\mu }k_{1}^{\nu
}k_{3}^{\lambda }k_{3}^{\gamma )}-s(\frac{s^{2}}{4}-1)(t+2)k_{1}^{(\mu
}k_{3}^{\nu }k_{3}^{\lambda }k_{3}^{\gamma )}  \notag \\
&&+(\frac{t^{2}}{4}-t)(\frac{t^{2}}{4}-1)k_{1}^{\mu }k_{1}^{\nu
}k_{1}^{\lambda }k_{1}^{\gamma }],  \TCItag{41}
\end{eqnarray}%
where $s=-(k_{1}+k_{2})^{2},t=-(k_{2}+k_{3})^{2},$ and $u$ =$%
-(k_{1}+k_{3})^{2}$are the Mandelstam variables. We have chosen the second
state to be the tensor and have done the $SL(2,R)$ gauge fixing and
restricted to the $s-t$ channel by setting $x_{1}=0,0\leq x_{2}\leq
1,x_{3}=1,x_{4}=\infty .$ One easily sees from eq.(41) that there are no
terms containing $\eta ^{\mu \nu }$on the right hand side of $\mathcal{T}%
_{3,1}^{\mu \nu \lambda \gamma }$. This is due to the normal ordering of the
tensor vertex operator  :$\partial x^{\mu }\partial x^{\nu }\partial
x^{\lambda }\partial x^{\gamma }e^{ik\cdot x}:$, and there is no
contribution of terms resulting from contraction within the tensor vertex
when doing the amplitude calculation. Thus the trace part of the rank-four
amplitude is fixed by the spin-four amplitude by the conformal extension
mentioned in the beginning of this section. That is, the rank-four amplitude 
$\mathcal{T}_{3,1}^{\mu \nu \lambda \gamma }$ is fixed by the spin-four
amplitude. This result can be easily generalized to N-point amplitudes
containing more than one tensor state.

Take a representative of the positive-norm scalar state at this mass level
to be \cite{13}

\begin{eqnarray}
&&[-(\eta _{\mu \nu }+\frac{13}{3}k_{\mu }k_{\nu })\alpha _{-2}^{\mu \nu
}+-i(\frac{20}{9}k_{\mu }k_{\nu }k_{\rho }+\frac{2}{3}k_{\mu }\eta _{\nu
\rho }+\frac{13}{3}k_{\rho }\eta _{\mu \nu })\alpha _{-1}^{\mu \nu }\alpha
_{-2}^{\rho }  \notag \\
&&+(\frac{23}{81}k_{\mu }k_{\nu }k_{\rho }k_{\sigma }+\frac{32}{27}k_{\mu
}k_{\nu }\eta _{\rho \sigma }+\frac{19}{18}\eta _{\mu \nu }\eta _{\rho
\sigma })\alpha _{-1}^{\mu \nu \rho \sigma }]\left\vert 0,k\right\rangle . 
\TCItag{42}
\end{eqnarray}%
It turns out that one can't gauge away the first term in eq.(42) by using
the gauge transformations induced by the two singlet zero-norm states as in
the case of positive-norm spin-two state. However, since the amplitude
corresponding to $\alpha _{-2}^{\mu \nu }$ has been fixed by those of two
higher spin states, we conclude that the positive-norm scalar state
amplitude is again fixed by those of two higher spin states. This concludes
the justification of decoupling conjecture for spin-four level. We stress
here that the mechanisms that is responsible for this decoupling is the
existence of two-types of zero-norm states and the factorization of stringy
vertex, which are both due to 2D infinite dimensional worldsheet conformal
symmetry.

The positive-norm states at level five were calculated in Ref \cite{12} to be

\begin{equation}
\epsilon _{\mu \nu \lambda \beta \gamma }\alpha _{-1}^{\mu \nu \lambda \beta
\gamma }\left\vert 0,k\right\rangle \text{ \ }\raisebox{0.06in}{\fbox{%
\rule[0.04cm]{0.04cm}{0cm}}}\raisebox{0.06in}{\fbox{%
\rule[0.04cm]{0.04cm}{0cm}}}\raisebox{0.06in}{\fbox{%
\rule[0.04cm]{0.04cm}{0cm}}}\raisebox{0.06in}{\fbox{%
\rule[0.04cm]{0.04cm}{0cm}}}\raisebox{0.06in}{\fbox{%
\rule[0.04cm]{0.04cm}{0cm}}},  \tag{43}
\end{equation}

\begin{equation}
\epsilon _{\mu \nu \lambda ,\beta }\alpha _{-1}^{\mu \nu \lambda }\alpha
_{-2}^{\beta }\left\vert 0,k\right\rangle \text{ \ }\raisebox{0.06in}{\fbox{%
\rule[0.04cm]{0.04cm}{0cm}}}\hspace{-0.094in}\hspace{-0.04cm}%
\raisebox{-.047in}{\fbox{\rule[0.04cm]{0.04cm}{0cm}}}\hspace{-0.006in}%
\hspace{-0.006in}\hspace{0.02cm}\raisebox{0.06in}{\fbox{%
\rule[0.04cm]{0.04cm}{0cm}}}\hspace{-0.025cm}\raisebox{0.06in}{\fbox{%
\rule[0.04cm]{0.04cm}{0cm}}},  \tag{44}
\end{equation}

\begin{equation}
\epsilon _{\mu ,\nu \lambda }(\alpha _{-1}^{\mu }\alpha _{-2}^{\nu \lambda }-%
\frac{4}{3}\alpha _{-1}^{\mu \nu }\alpha _{-3}^{\lambda })\left\vert
0,k\right\rangle \text{ \ }\raisebox{0.06in}{\fbox{%
\rule[0.04cm]{0.04cm}{0cm}}}\hspace{-0.094in}\hspace{-0.04cm}%
\raisebox{-.047in}{\fbox{\rule[0.04cm]{0.04cm}{0cm}}}\hspace{-0.006in}%
\hspace{-0.006in}\hspace{0.02cm}\raisebox{0.06in}{\fbox{%
\rule[0.04cm]{0.04cm}{0cm}}},  \tag{45}
\end{equation}

\begin{equation}
\lbrack \frac{4}{5!(D+5)}\epsilon _{\mu \nu \lambda }\eta _{\beta \gamma
}^{T}\alpha _{-1}^{\mu \nu \lambda \beta \gamma }+\epsilon _{\mu \nu \lambda
}(\alpha _{-1}^{\mu }\alpha _{-2}^{\nu \lambda }-\frac{4}{3}\alpha
_{-1}^{\mu \nu }\alpha _{-3}^{\lambda })]\left\vert 0,k\right\rangle \text{
\ }\raisebox{0.06in}{\fbox{\rule[0.04cm]{0.04cm}{0cm}}}\raisebox{0.06in}{%
\fbox{\rule[0.04cm]{0.04cm}{0cm}}}\raisebox{0.06in}{\fbox{%
\rule[0.04cm]{0.04cm}{0cm}}},  \tag{46}
\end{equation}

\begin{equation}
\lbrack \frac{5}{6(D+1)}\eta _{(\mu \nu }^{T}\epsilon _{\lambda )\beta
}\alpha _{-1}^{\mu \nu \lambda }\alpha _{-2}^{\beta }+\epsilon _{\mu \nu
}(\alpha _{-2}^{\mu }\alpha _{-3}^{\nu }-\frac{1}{2}\alpha _{-1}^{\mu
}\alpha _{-4}^{\nu })]\left\vert 0,k\right\rangle \text{  }%
\raisebox{0.06in}{\fbox{\rule[0.04cm]{0.04cm}{0cm}}}\hspace{-0.094in}\hspace{%
-0.04cm}\raisebox{-.047in}{\fbox{\rule[0.04cm]{0.04cm}{0cm}}}\text{\ }, 
\tag{47}
\end{equation}

and

\begin{eqnarray}
&&[\frac{D-2}{80(D+3)}\eta _{(\mu \nu }^{T}\eta _{\lambda \beta
}^{T}\epsilon _{\gamma )}\alpha _{-1}^{\mu \nu \lambda \beta \gamma }+(\eta
_{\mu \nu }^{T}\epsilon _{\lambda }-\frac{1}{2}(D-1)\epsilon _{(\mu }\eta
_{\nu )\lambda }^{T})\alpha _{-1}^{\mu \nu }\alpha _{-3}^{\lambda } 
\TCItag{48} \\
&&+\frac{3}{4}(D\epsilon _{\mu }\eta _{\nu \lambda }^{T}-\eta _{\mu (\nu
}^{T}\epsilon _{\lambda )})\alpha _{-1}^{\mu }\alpha _{-2}^{\nu \lambda
})]\left\vert 0,k\right\rangle \text{ }\raisebox{0.06in}{\fbox{%
\rule[0.04cm]{0.04cm}{0cm}}}.  \notag
\end{eqnarray}%
According to our decoupling conjecture, states (46), (47) and (48) should be
decoupled. Note that states (27) and (45) are different in the $\alpha
_{i}^{\prime }s$ operator content although they share the same Young
diagram. One corresponds to $\alpha _{-1}^{\mu }\alpha _{-2}^{\nu \lambda }$
and the other $\alpha _{-1}^{\mu \nu }\alpha _{-3}^{\lambda }$ or vice
versa. With the explicit form of zero-norm states calculated in section III,
we can now justify the decoupling conjecture at level five. The terms $%
\alpha _{-1}^{(\mu }\alpha _{-2}^{\nu \lambda )}$ and $\alpha _{-1}^{(\mu
\nu }\alpha _{-3}^{\lambda )}$ in eq.(46) can be gauged away by zero-norm
states in eqs.(25) and (26), and the amplitude corresponding to $\alpha
_{-1}^{(\mu \nu \lambda }\alpha _{-2}^{\beta )}$ is fixed by that of $\alpha
_{-1}^{\mu \nu \lambda \beta \gamma }$ through zero-norm state in eq.(24)
and our observation discussed in the beginning of this section. Thus the
amplitude of state (46) is fixed by those of states (43) and (44). Now turn
to state (47). The terms $\alpha _{-2}^{[\mu }\alpha _{-3}^{\nu ]}$ and $%
\alpha _{-1}^{[\mu }\alpha _{-4}^{\nu ]}$ can be gauged away by zero-norm
states in eqs.(28) and (29), the amplitudes corresponding to $\alpha
_{-1}^{(\mu }\alpha _{-2}^{\nu \lambda )}$ and $\alpha _{-1}^{(\mu \nu
}\alpha _{-3}^{\lambda )}$ are fixed by those of states in eqs.(43) and (44)
through zero-norm states in eqs.(25) and (26). Finally the amplitude of
mixed-symmetric $\alpha _{-1}^{\mu }\alpha _{-2}^{\nu \lambda }$ (or $\alpha
_{-1}^{\mu \nu }\alpha _{-3}^{\lambda }$) is fixed by those of states (43),
(44) and (45). Thus the amplitude of state (47) is fixed by those of states
(43), (44) and (45). Similar analysis shows that the amplitude of state (48)
is again fixed by those of states (43), (44) and (45). This completes the
justification of our decoupling conjecture at level five.

The decoupling calculation presented in this paper by the S-matrix approach
can be easily generalized to the closed string theory by making use of the
simple relation between closed and open string amplitudes in Ref \cite{17}.
A similar generalization to the closed string theory can also be done for
the massive worldsheet sigma-model approach. Our calculation in this section
justifies two previous independent calculations based on the massive
worldsheet sigma-model approach \cite{3} and WSFT approach \cite{6}.

\section{Acknowledgments}

I would like to thank Physics Departments of National Taiwan University and
Simon-Fraser University, where part of this work was completed during my
sabbatical visits. I thank Chuan-Tsung Chan and Pei-Ming Ho for many
valuable discussions. This work is supported in part by a grant of National
Science Council and a travelling fund of government of Taiwan.

\appendix

\section{ \ \ \ \ \ \ }

The Young tabulations of all physical states solutions of eq.(3) up to level
six, including two types of zero-norm state solutions of eqs.(5) and (6),
are listed in the following table

\begin{tabular}{|l|l|l|}
\hline
massive level & positive-norm states & zero-norm states \\ \hline
$M^{2}=-2$ & $\bullet $ &  \\ \hline
$M^{2}=0$ & $\raisebox{0.06in}{\fbox{\rule[0.04cm]{0.04cm}{0cm}}}$ & $%
\bullet $ (singlet) \\ \hline
$M^{2}=2$ & $\raisebox{0.06in}{\fbox{\rule[0.04cm]{0.04cm}{0cm}}}%
\raisebox{0.06in}{\fbox{\rule[0.04cm]{0.04cm}{0cm}}}$ & $\raisebox{0.06in}{%
\fbox{\rule[0.04cm]{0.04cm}{0cm}}},$ $\bullet $ \\ \hline
$M^{2}=4$ & $\raisebox{0.06in}{\fbox{\rule[0.04cm]{0.04cm}{0cm}}}%
\raisebox{0.06in}{\fbox{\rule[0.04cm]{0.04cm}{0cm}}}\raisebox{0.06in}{\fbox{%
\rule[0.04cm]{0.04cm}{0cm}}},\raisebox{0.06in}{\fbox{%
\rule[0.04cm]{0.04cm}{0cm}}}\hspace{-0.094in}\hspace{-0.04cm}%
\raisebox{-.047in}{\fbox{\rule[0.04cm]{0.04cm}{0cm}}}$ & $%
\raisebox{0.06in}{\fbox{\rule[0.04cm]{0.04cm}{0cm}}}\raisebox{0.06in}{\fbox{%
\rule[0.04cm]{0.04cm}{0cm}}},2\times \raisebox{0.06in}{\fbox{%
\rule[0.04cm]{0.04cm}{0cm}}},\bullet $ \\ \hline
$M^{2}=6$ & $\raisebox{0.06in}{\fbox{\rule[0.04cm]{0.04cm}{0cm}}}%
\raisebox{0.06in}{\fbox{\rule[0.04cm]{0.04cm}{0cm}}}\raisebox{0.06in}{\fbox{%
\rule[0.04cm]{0.04cm}{0cm}}}\raisebox{0.06in}{\fbox{%
\rule[0.04cm]{0.04cm}{0cm}}},\raisebox{0.06in}{\fbox{%
\rule[0.04cm]{0.04cm}{0cm}}}\hspace{-0.094in}\hspace{-0.04cm}%
\raisebox{-.047in}{\fbox{\rule[0.04cm]{0.04cm}{0cm}}}\hspace{-0.006in}%
\hspace{-0.006in}\hspace{0.02cm}\raisebox{0.06in}{\fbox{%
\rule[0.04cm]{0.04cm}{0cm}}},\raisebox{0.06in}{\fbox{%
\rule[0.04cm]{0.04cm}{0cm}}}\raisebox{0.06in}{\fbox{%
\rule[0.04cm]{0.04cm}{0cm}}},\bullet $ & $\raisebox{0.06in}{\fbox{%
\rule[0.04cm]{0.04cm}{0cm}}}\raisebox{0.06in}{\fbox{%
\rule[0.04cm]{0.04cm}{0cm}}}\raisebox{0.06in}{\fbox{%
\rule[0.04cm]{0.04cm}{0cm}}},\raisebox{0.06in}{\fbox{%
\rule[0.04cm]{0.04cm}{0cm}}}\hspace{-0.094in}\hspace{-0.04cm}%
\raisebox{-.047in}{\fbox{\rule[0.04cm]{0.04cm}{0cm}}},2\times %
\raisebox{0.06in}{\fbox{\rule[0.04cm]{0.04cm}{0cm}}}\raisebox{0.06in}{\fbox{%
\rule[0.04cm]{0.04cm}{0cm}}},3\times \raisebox{0.06in}{\fbox{%
\rule[0.04cm]{0.04cm}{0cm}}},$ $2\times \bullet $ \\ \hline
$M^{2}=8$ & $\raisebox{0.06in}{\fbox{\rule[0.04cm]{0.04cm}{0cm}}}%
\raisebox{0.06in}{\fbox{\rule[0.04cm]{0.04cm}{0cm}}}\raisebox{0.06in}{\fbox{%
\rule[0.04cm]{0.04cm}{0cm}}}\raisebox{0.06in}{\fbox{%
\rule[0.04cm]{0.04cm}{0cm}}}\raisebox{0.06in}{\fbox{%
\rule[0.04cm]{0.04cm}{0cm}}},\raisebox{0.06in}{\fbox{%
\rule[0.04cm]{0.04cm}{0cm}}}\hspace{-0.094in}\hspace{-0.04cm}%
\raisebox{-.047in}{\fbox{\rule[0.04cm]{0.04cm}{0cm}}}\hspace{-0.006in}%
\hspace{-0.006in}\hspace{0.02cm}\raisebox{0.06in}{\fbox{%
\rule[0.04cm]{0.04cm}{0cm}}}\raisebox{0.06in}{\fbox{%
\rule[0.04cm]{0.04cm}{0cm}}},\raisebox{0.06in}{\fbox{%
\rule[0.04cm]{0.04cm}{0cm}}}\hspace{-0.094in}\hspace{-0.04cm}%
\raisebox{-.047in}{\fbox{\rule[0.04cm]{0.04cm}{0cm}}}\hspace{-0.006in}%
\hspace{-0.006in}\hspace{0.02cm}\raisebox{0.06in}{\fbox{%
\rule[0.04cm]{0.04cm}{0cm}}},\raisebox{0.06in}{\fbox{%
\rule[0.04cm]{0.04cm}{0cm}}}\raisebox{0.06in}{\fbox{%
\rule[0.04cm]{0.04cm}{0cm}}}\raisebox{0.06in}{\fbox{%
\rule[0.04cm]{0.04cm}{0cm}}},\raisebox{0.06in}{\fbox{%
\rule[0.04cm]{0.04cm}{0cm}}}\hspace{-0.094in}\hspace{-0.04cm}%
\raisebox{-.047in}{\fbox{\rule[0.04cm]{0.04cm}{0cm}}},\raisebox{0.06in}{%
\fbox{\rule[0.04cm]{0.04cm}{0cm}}}$ & $\raisebox{0.06in}{\fbox{%
\rule[0.04cm]{0.04cm}{0cm}}}\raisebox{0.06in}{\fbox{%
\rule[0.04cm]{0.04cm}{0cm}}}\raisebox{0.06in}{\fbox{%
\rule[0.04cm]{0.04cm}{0cm}}}\raisebox{0.06in}{\fbox{%
\rule[0.04cm]{0.04cm}{0cm}}},\raisebox{0.06in}{\fbox{%
\rule[0.04cm]{0.04cm}{0cm}}}\hspace{-0.094in}\hspace{-0.04cm}%
\raisebox{-.047in}{\fbox{\rule[0.04cm]{0.04cm}{0cm}}}\hspace{-0.006in}%
\hspace{-0.006in}\hspace{0.02cm}\raisebox{0.06in}{\fbox{%
\rule[0.04cm]{0.04cm}{0cm}}},2\times \raisebox{0.06in}{\fbox{%
\rule[0.04cm]{0.04cm}{0cm}}}\raisebox{0.06in}{\fbox{%
\rule[0.04cm]{0.04cm}{0cm}}}\raisebox{0.06in}{\fbox{%
\rule[0.04cm]{0.04cm}{0cm}}},2\times \raisebox{0.06in}{\fbox{%
\rule[0.04cm]{0.04cm}{0cm}}}\hspace{-0.094in}\hspace{-0.04cm}%
\raisebox{-.047in}{\fbox{\rule[0.04cm]{0.04cm}{0cm}}},4\times %
\raisebox{0.06in}{\fbox{\rule[0.04cm]{0.04cm}{0cm}}}\raisebox{0.06in}{\fbox{%
\rule[0.04cm]{0.04cm}{0cm}}},5\times \raisebox{0.06in}{\fbox{%
\rule[0.04cm]{0.04cm}{0cm}}},3\times \bullet $ \\ \hline
$M^{2}=10$ & $%
\begin{array}{c}
\raisebox{0.06in}{\fbox{\rule[0.04cm]{0.04cm}{0cm}}}\raisebox{0.06in}{\fbox{%
\rule[0.04cm]{0.04cm}{0cm}}}\raisebox{0.06in}{\fbox{%
\rule[0.04cm]{0.04cm}{0cm}}}\raisebox{0.06in}{\fbox{%
\rule[0.04cm]{0.04cm}{0cm}}}\raisebox{0.06in}{\fbox{%
\rule[0.04cm]{0.04cm}{0cm}}}\raisebox{0.06in}{\fbox{%
\rule[0.04cm]{0.04cm}{0cm}}},\raisebox{0.06in}{\fbox{%
\rule[0.04cm]{0.04cm}{0cm}}}\hspace{-0.094in}\hspace{-0.04cm}%
\raisebox{-.047in}{\fbox{\rule[0.04cm]{0.04cm}{0cm}}}\hspace{-0.006in}%
\hspace{-0.006in}\hspace{0.02cm}\raisebox{0.06in}{\fbox{%
\rule[0.04cm]{0.04cm}{0cm}}}\raisebox{0.06in}{\fbox{%
\rule[0.04cm]{0.04cm}{0cm}}}\raisebox{0.06in}{\fbox{%
\rule[0.04cm]{0.04cm}{0cm}}},\raisebox{0.06in}{\fbox{%
\rule[0.04cm]{0.04cm}{0cm}}}\raisebox{0.06in}{\fbox{%
\rule[0.04cm]{0.04cm}{0cm}}}\raisebox{0.06in}{\fbox{%
\rule[0.04cm]{0.04cm}{0cm}}}\raisebox{0.06in}{\fbox{%
\rule[0.04cm]{0.04cm}{0cm}}},\raisebox{0.06in}{\fbox{%
\rule[0.04cm]{0.04cm}{0cm}}}\hspace{-0.094in}\hspace{-0.04cm}%
\raisebox{-.047in}{\fbox{\rule[0.04cm]{0.04cm}{0cm}}}\hspace{-0.006in}%
\hspace{-0.006in}\hspace{0.02cm}\raisebox{0.06in}{\fbox{%
\rule[0.04cm]{0.04cm}{0cm}}}\raisebox{0.06in}{\fbox{%
\rule[0.04cm]{0.04cm}{0cm}}} \\ 
\raisebox{0.06in}{\fbox{\rule[0.04cm]{0.04cm}{0cm}}}\hspace{-0.094in}\hspace{%
-0.04cm}\raisebox{-.047in}{\fbox{\rule[0.04cm]{0.04cm}{0cm}}}%
\raisebox{0.06in}{\fbox{\rule[0.04cm]{0.04cm}{0cm}}}\hspace{-0.094in}\hspace{%
-0.04cm}\raisebox{-.047in}{\fbox{\rule[0.04cm]{0.04cm}{0cm}}},%
\raisebox{0.06in}{\fbox{\rule[0.04cm]{0.04cm}{0cm}}}\raisebox{0.06in}{\fbox{%
\rule[0.04cm]{0.04cm}{0cm}}}\raisebox{0.06in}{\fbox{%
\rule[0.04cm]{0.04cm}{0cm}}},\raisebox{0.06in}{\fbox{%
\rule[0.04cm]{0.04cm}{0cm}}}\hspace{-0.094in}\hspace{-0.04cm}%
\raisebox{-.047in}{\fbox{\rule[0.04cm]{0.04cm}{0cm}}}\hspace{-0.006in}%
\hspace{-0.006in}\hspace{0.02cm}\raisebox{0.06in}{\fbox{%
\rule[0.04cm]{0.04cm}{0cm}}},\raisebox{0.06in}{\fbox{%
\rule[0.04cm]{0.04cm}{0cm}}}\hspace{-0.094in}\hspace{-0.04cm}%
\raisebox{-.047in}{\fbox{\rule[0.04cm]{0.04cm}{0cm}}}\hspace{-0.28cm}%
\raisebox{-.150in}{\fbox{\rule[0.04cm]{0.04cm}{0cm}}},\text{\ \ }\hspace{%
-0.094in}2\times \raisebox{0.06in}{\fbox{\rule[0.04cm]{0.04cm}{0cm}}}%
\raisebox{0.06in}{\fbox{\rule[0.04cm]{0.04cm}{0cm}}},\raisebox{0.06in}{%
\fbox{\rule[0.04cm]{0.04cm}{0cm}}},\bullet%
\end{array}%
$ & $%
\begin{array}{c}
\raisebox{0.06in}{\fbox{\rule[0.04cm]{0.04cm}{0cm}}}\raisebox{0.06in}{\fbox{%
\rule[0.04cm]{0.04cm}{0cm}}}\raisebox{0.06in}{\fbox{%
\rule[0.04cm]{0.04cm}{0cm}}}\raisebox{0.06in}{\fbox{%
\rule[0.04cm]{0.04cm}{0cm}}}\raisebox{0.06in}{\fbox{%
\rule[0.04cm]{0.04cm}{0cm}}},2\times \raisebox{0.06in}{\fbox{%
\rule[0.04cm]{0.04cm}{0cm}}}\raisebox{0.06in}{\fbox{%
\rule[0.04cm]{0.04cm}{0cm}}}\raisebox{0.06in}{\fbox{%
\rule[0.04cm]{0.04cm}{0cm}}}\raisebox{0.06in}{\fbox{%
\rule[0.04cm]{0.04cm}{0cm}}},\raisebox{0.06in}{\fbox{%
\rule[0.04cm]{0.04cm}{0cm}}}\hspace{-0.094in}\hspace{-0.04cm}%
\raisebox{-.047in}{\fbox{\rule[0.04cm]{0.04cm}{0cm}}}\hspace{-0.006in}%
\hspace{-0.006in}\hspace{0.02cm}\raisebox{0.06in}{\fbox{%
\rule[0.04cm]{0.04cm}{0cm}}}\raisebox{0.06in}{\fbox{%
\rule[0.04cm]{0.04cm}{0cm}}},3\times \raisebox{0.06in}{\fbox{%
\rule[0.04cm]{0.04cm}{0cm}}}\hspace{-0.094in}\hspace{-0.04cm}%
\raisebox{-.047in}{\fbox{\rule[0.04cm]{0.04cm}{0cm}}}\hspace{-0.006in}%
\hspace{-0.006in}\hspace{0.02cm}\raisebox{0.06in}{\fbox{%
\rule[0.04cm]{0.04cm}{0cm}}}, \\ 
4\times \raisebox{0.06in}{\fbox{\rule[0.04cm]{0.04cm}{0cm}}}%
\raisebox{0.06in}{\fbox{\rule[0.04cm]{0.04cm}{0cm}}}\raisebox{0.06in}{\fbox{%
\rule[0.04cm]{0.04cm}{0cm}}},4\times \raisebox{0.06in}{\fbox{%
\rule[0.04cm]{0.04cm}{0cm}}}\hspace{-0.094in}\hspace{-0.04cm}%
\raisebox{-.047in}{\fbox{\rule[0.04cm]{0.04cm}{0cm}}},7\times %
\raisebox{0.06in}{\fbox{\rule[0.04cm]{0.04cm}{0cm}}}\raisebox{0.06in}{\fbox{%
\rule[0.04cm]{0.04cm}{0cm}}},8\times \raisebox{0.06in}{\fbox{%
\rule[0.04cm]{0.04cm}{0cm}}},6\times \bullet%
\end{array}%
$ \\ \hline
\end{tabular}

Note that the Young tabulations of zero-norm states at level n are subset of
the sum of all physical states at levels n-1 and n-2.

\end{document}